\documentclass[preprint,aps,epsfig]{revtex4}
\usepackage{graphics,psfig,graphicx}
\usepackage{amsbsy}
\newcommand {\be}{\begin{equation}}
\newcommand {\ee}{\end{equation}}
\newcommand {\bea}{\begin{eqnarray}}
\newcommand {\ea}{\end{eqnarray*}}
\newcommand {\ba}{\begin{eqnarray*}}
\newcommand {\eea}{\end{eqnarray}}

\newcommand {\lleb} {{\mathcal L}}

\newcommand {\bra}{\langle}
\newcommand {\ket}{\rangle}

\newcommand {\refeq}[1] {(\ref{#1})}
\newcommand {\rdd}{\sqrt{\bra R^2 \ket}}
\begin{document}
\title{Comparative study of Rare Gas-H$_2$ triatomic complexes.}

\author{Paolo Barletta}
\email{p.barletta@ucl.ac.uk}
\affiliation{Department of Physics and Astronomy, University College London, 
Gower Street, London WC1E 6BT, United Kingdom}

\date{\today}


\begin{abstract}
This paper presents a comparative analysis of complexes made of one Rare Gas (Rg) atom and molecular hydrogen, for all five stable Rg atoms. In particular, the vibrational band origins have been calculated, as well as the structural properties of the associated wavefunctions. The study is extended to cold Rg-H$_2$ scattering. The molecular systems are studied variationally using a very simple, yet effective, trial wavefunction. A large number of Potential Energy Surfaces available from the literature is considered. A comparative analysis shows that differences of up to two orders of magnitude exist for the zero energy elastic cross sections of the five complexes. Corrections to the model have also been considered, showing no significant effect.
\end{abstract}

\maketitle


\section{Introduction}

Over the last decades, complexes made of one rare gas atom and molecular hydrogen (Rg-H$_2$) have provided an optimum testing ground for {\it ab initio} and experimental techniques \cite{revbfh88}. A  wealth of data for the Rg-H$_2$ complexes has been collected in a large number of very accurate experimental investigations. This includes high-precision measurements of the rotational-vibrational transitions (mostly in the microwave and infrared regions), elastic and inelastic differential cross-sections, and virial coefficients data. In parallel, these systems have stimulated a remarkable interest in the theoretical community, providing the possibility of carefully testing {\it ab initio} methods. For example, ArH$_2$ has become a popular benchmark for theory, especially for calculations of rotational and vibrational pre-dissociation effects \cite{expmck96}, as for this system the widest range of spectroscopic, collisional and bulk property data are available \cite{arh26}.  \\

In recent years the advent of Bose-Einstein condensates of atoms and molecules has revitalised the interest in van der Waals complexes. The areas of cold and ultra-cold physics \cite{revdfk04} ares rapidly evolving and attracting a considerable interest in the scientific community. This is due to the possibility of cooling atoms and molecules at temperatures below 1 K, which creates the capability of experimentally controlling chemical reactions, allowing accurate studies of ro-vibrational energy transfers and coherent chemistry.  A rather promising experimental route being pursued is the creation of ultra-cold molecules by means of sympathetic cooling with other ultracold species. Consequently, a number of groups have begun the investigation of ultra-cold collisions and chemical reactions from the theoretical point of view \cite{revhs06}, calculating {\it ab initio} elastic and inelastic cross-section, branching ratios, resonance energies and widths, etc. In particular, the calculation of cross sections (both elastic and inelastic) between different species is hugely importance for the planning and realisation of experiments with ultra-cold species. \\

The main ingredient in an {\it ab initio} calculation is the Potential Energy Surface (PES) modelling the interaction among the atoms, within the Born-Oppenheimer approximation. In recent years, a large number of PESs have been produced for the five stable Rg-H$_2$ complexes (Rg = He, Ne, Ar, Kr, and Xe). Initially, the most accurate PESs were produced semi-empirically, by means of fits to measured data \cite{revbfh88}. With the advances in computational power and quantum chemistry methods, purely {\it ab initio} PESs of comparable accuracy are now available. In particular, extensive work has been done on Ar-H$_2$ \cite{arh27,arh22,arh26,arh23}, which is one of the most thoroughly studied atom-diatom complexes. More recently, the He-H$_2$ PES has attracted a considerable attention. Due to the weakness of the He-H$_2$ interaction, the existence of a bound state for this complex has long been debated.  Recently, Kalinin {\it et al.} have produced evidence for its existence by means of a transmission grating diffraction experiment \cite{hekkr04}. In parallel, a fully {\it ab initio} study of its He-H$_2$ PES has been  carried with high accuracy \cite{hebmp03}, which has inspired a number of theoretical papers on the vibrational spectrum of the complex. \\ 

Despite the very large availability of PESs, relatively few groups have reported accurate quantum mechanical calculations for Rg-H$_2$ complexes. The He-H$_2$ system has been object of three theoretical papers \cite{heyll99,heggl05,hexp06} aiming to assess the properties of its halo state, following Kalinin {\it et al.}'s experiment, whereas He-H$_2$ collision processes at low temperature were studied by other groups \cite{hebfd98,helrm05}. The ro-vibrational spectrum of Ar$_2$ was calculated variationally by Moszynski and co-workers \cite{mismjw94}, whereas H$_2$ and D$_2$ collisions with Ar atoms were investigated by Uudus and co-workers \cite{arh24}. The rotational-vibrational spectrum of Kr-H$_2$ was analysed by Wei and co-workers \cite{krh21} and by Zhou and Xie \cite{krh22}. The former have produced a highly accurate semi-empirical PES, also performing all dimensional calculations of the molecule's vibrational-rotational levels, whereas the latter reported a purely {\it ab initio} PES calculated at the CCSD(T) level of theory together with the calculation of the associated vibrational-rotational levels.  \\

The object of this work is a comparative analysis of Rg-H$_2$ complexes, using, where possible, more than a single PES for each molecule. Given the very large number of PESs available in literature, our aim is not to provide a fully comprehensive study of them all. We chose to selecting some of the most popular semi-empirical PESs from the 1980s, such as those of Rodwell and Scoles \cite{nersc82}, and Le Roy and Hutson \cite{xerh86}, and all the most recent ones, including the one by Boothroyd {\it et al.} \cite{hebmp03} for He-H$_2$, Bissonnette {\it et al.} \cite{arh26} for Ar-H$_2$, and a few others. Although some of the surfaces are known in literature with proper names, each of them were assigned a new acronym in this paper in order to uniform the notation among the different complexes. The molecules are studied by variational methods, by using the Rayleigh-Ritz and Kohn principle for bound and continuum states, respectively. The trial wavefunction used is based on a simple, but effective model, which assumes that the Rg atom rotates/vibrates around a the H$_2$ molecule stiff in its vibrational-rotational ground state. This model has been used in variational many-body calculations \cite{misbw92}, and Gianturco {\it et al.} proved it sound also in comparison to very accurate full dimensional calculations for the halo He-H$_2$ state \cite{heggl05}. We extend their work to also study cold scattering, with applications to the five different Rg-H$_2$ complexes. \\

This paper is organised as follows. The next Section will briefly present the model used to study the Rg-H$_2$ systems, with details of the computational procedure. In Section III the results obtained for the five complexes are presented and discussed individually. The last Section is devoted to the comparative analysis of the different complexes, as well as the conclusions. This Section presents the elastic cross-sections at cold energy, as well as the structural and energetic properties of the vibrational molecular states.

\section{Computational details}

The H$_2$ molecule is strongly bound compared to RgH$_2$ binding. Using the H-H potential from Ref. \cite{dimsch88}, the lowest rotational levels in the ground vibrational band result to be 118.4 cm$^{-1}$, 354.2 cm$^{-1}$ and 705.1 cm$^{-1}$ for the states with $\ell=1$, $\ell=2$ and $\ell=3$, respectively, where the energy of the ground state ($\ell=0$) is taken as zero energy point.  In this type of problems, it is common practice to write the system wavefunction in terms of the diatom molecular states \cite{misscp00}
\be
\Psi =  \sum_{\nu \ell \ell_R} \phi_{\nu \eta}(r) f_{\nu \eta}(R) \left[ Y_\ell(\hat{r}) Y_\eta(\hat{R}) \right]_{LM}, 
\label{th}
\ee
where $\phi_{\nu\ell}(r)$ is the radial part of the H$_2$ wavefunction with $\nu$ and $\ell$ its vibrational and rotational quantum numbers, respectively, $Y_\ell(\hat{r})$ and $ Y_\eta(\hat{R})$ are spherical harmonics associated to the Jacobi vectors $r$ and $R$, respectively, and $f_{\nu \eta}(R)$ is a set of unknown functions to be determined. The molecule is considered in a state of total angular momentum $L$, and $z$-axis component $M$. Due to the weakness of the Rg-H$_2$ potential we can restrict our analysis to $L=0$. The heavier Rg atoms form compounds with $L>0$ but we will ignore them for the purposes of this work. For the particular case of $L=0$ we have that $\ell=\eta$  and the dependence of $\Psi$ on the four polar angles $\hat{r}$ and $\hat{R}$ can be reduced to just one non-trivial coordinate by integrating out the Euler angles 
\be
\left[ Y_\ell(\hat{r}) Y_\ell(\hat{R}) \right]_{00} = P_\ell(\mu)
\ee
where $\mu=\hat(r)\cdot\hat(R)$ and $P_\ell $ is a Legendre polynomial. \\

 At this point it is possible to notice that the expansion of eq. \ref{th} involves H$_2$ states with high internal energy if compared to the Rg-H$_2$ interaction. In fact, the vibrational energies of the Rg-H$_2$ complexes are 10-100 times smaller (vide infra). It is reasonable to expect that the presence of the Rg atom will only slightly affect the H$_2$ diatom, even a short distances. The Rg-H$_2$ complex can thus be reasonably modelled, at low temperatures, by assuming that the Rg-H$_2$ potential only slightly perturbs the H-H Hamiltonian. Under this assumption, the two H$_2$ quantum numbers $\nu$ and $\ell$ can be considered good quantum numbers for the three-body system. In mathematical terms, this corresponds to retaining only the first term of the expansion of eq. \ref{th}. A presumably very accurate trial wavefunction for those complexes can thus be written as
\be 
\Psi =  \phi_{00}(r) f(R)/R \, P_0(\mu) , 
\label{trial}
\ee
where $f(R)$ is an unknown function to be determined. \\

 A very close approach was already shown to be surprisingly accurate by Gianturco {\it et al.} \cite{heggl05}, with an application to the binding energy of the HeH$_2$ system. In this work, we are interested in studying how well the five RgH$_2$ complexes can be modelled by using the trial wavefunction of eq. \refeq{trial}. We are interested in calculating not only the vibrational energies, but also the the associated wavefunctions and their structural properties. Furthermore, we can apply the same trial wavefunction to study the continuum energy region, in order to calculate the atom-dimer scattering length, the effective range, the phase-shift over a range of energies, and finally the Rg-H$_2$ cross sections. Those calculations will finally allow us to make a comparative analysis of the scattering properties in the five different complexes and of the several available PESs. \\

A general expression for the Rg-H$_2$ intermolecular PES is the following:
\be
V(r,R,\mu) = V_{\rm Rg-H_2}(r,R,\mu) + V_{\rm HH}(r), 
\ee
where $V_{\rm HH}$ is the H-H potential when the Rg atom sits at an infinite distance from H$_2$, and $V_{\rm Rg-H_2}$ parameterises the weak interaction between H$_2$ and the Rg atom. The oldest potentials used in this work assumed a stiff H$_2$, and are expressed in terms of $R$ and $\mu$ only \cite{revbfh88}. \\

In eq.  \refeq{trial} the only unknown is the function $f(R)$ associated to the motion of the Rg atom with respect to H$_2$. If the variational principle is applied to the trial wavefunction $\Psi$, one is led to solving the one-dimensional equation 
\be
\left[ - \frac{\hbar^2}{2 \mu} \frac{d^2}{d R^2} + V_{eff}(R) - E \right] f(R) = 0 , 
\label{effeq}
\ee
where 
\be
V_{eff} = \int \phi_{00}^2(r) P_0^2(\mu) V(r,R,\mu) d r d\mu , 
\ee
and $E$ is the energy of the three-body system with respect to the H$_2$ binding energy. In particular, the zero point for the energy axis is chosen with the Rg atom at an infinite distance from the H$_2$ molecule in its vibrational-rotational ground state. All energies reported will use this convention. We will also use cm$^{-1}$ as units for energies and \AA for distances.  Equation \refeq{effeq} is solved using the Numerov algorithm, and the effective potential is calculated on a uniform grid of 10,000 points equally distributed between 0.01 and 100 \AA. The equation above can be solved for negative values of E, yielding upper bounds to the vibrational energies, and for positive values of E, yielding the complex scattering wavefunction $f_E$. In this latter case, a first order estimate for the observable of interest, say $\lleb$, is obtained from the asymptotic part of the function $f_E(R)$, and the second order estimate by means of the Kohn variational principle:
\be
\lleb^{2nd} = \lleb^{1st} - \bra f_E | \left. - \frac{\hbar^2}{2 \mu} \frac{d^2}{d R^2} + V_{eff}(R) - E \right.   | f_E \ket.
\ee

With the procedure just described the phase-shift $\delta(E)$ can be obtained over a range of energies E. The centre of mass elastic cross section is then expressed as
\be
\sigma(E) = 4\pi \frac{\sin^2 \delta}{k^2} , 
\ee
where $k=\sqrt{2\mu E/\hbar^2}$. At low energy the effective range expansion for $\delta$ reads \cite{taylor}
\be
\frac{\tan \delta}{k} = -\frac{1}{a_s} + k^2 \frac{r_{eff}}{2} , 
\ee
where $a_s$ is the scattering length and $r_{eff}$ the effective range. Consequently, the limit for elastic cross section at zero energy is
\be
\sigma(E=0) = 4\pi a_s^2 .
\ee
The effective range $r_{eff}$ is obtained by means of the integral relation \cite{taylor}
\be
r_{eff} = \frac{2}{a_s^2}\int \left[ f_0(R)^2-(R-a_s)^2 \right] dR ,
\ee
where $f_0(R)$ is the solution of eq. \refeq{effeq} for $E=0$, with the appropriate normalisation.\\

In order to test whether the ansatz in eq. \refeq{trial} for the wavefunction is a good choice, it is possible to calculate the contribution of the next H$_2$ channel in energy to the desired observables. Namely, a two channel trial wavefunction which includes the H$_2$ ($\nu=0,\ell=2$) state can be build as follows:
\be 
\Psi = \phi_{00}(r) f(R)/R \, P_0(\mu) + \phi_{02}(r) g(R)/R \, P_2(\mu), 
\label{trial2}
\ee
where $f$ and $g$ are two unknown functions to be determined . This choice is motivated by energy and symmetry considerations, as the state considered is the allowed H$_2$ rotational-vibrational state closest in energy to the the ground state, as the state with $\ell=1$ belongs to a different symmetry block. The implementation of the variational principle leads to a system of two one-dimensional coupled differential equations. In order to determine $f$ and $g$, they are expanded in a polynomial basis, and the coefficients of the expansion are determined by solving a generalised eigenvalue problem for the bound state or a linear problem for the scattering states. Results associated with the trial wavefunction \ref{trial2} will only be presented for a specific case, in order to show that the contribution of the state $(\nu=0,\ell=2)$ is very small.

\section{Results}

\subsection{He-H$_2$}

The He-H$_2$ system is the computationally most interesting one due to the presence of a halo state. Halo states \cite{misjrf04} are very weakly bound and extremely sensitive to the minimum details of the PES. As the He-H$_2$ van der Waals interaction is a consequence of big cancellations between the different terms of the molecular Hamiltonian, very accurate calculations of the PES are needed in order to provide a reliable {\it ab initio} PES and a subsequent estimate of the HeH$_2$ binding energy. The PES of Bissonnette {\it et al} \cite{hebmp03} (heh2c) was computed fitting a very large number of {\it ab initio} energies calculated at the MR-CISD (multiple-reference single and double excitation configuration interaction) level of theory. Other available PESs for this complex are the semi-empirical from Ref. \cite{nersc82} (heh2s), and from Ref. \cite{hemr94} (heh2b). Table \ref{tab1} shows calculated observable for the HeH$_2$ complex, namely the parameters $R_{min}$, $\epsilon$ and $R_0$, representing the position of the minimum and the well depth of the effective potential $V_{eff}$, and the point where it crosses the $R-$axis, respectively. The HeH$_2$ system has only one bound state, of energy $E$. Other listed quantities in Table \ref{tab1} are the mean values of the kinetic and potential energy, $\bra K \ket$ and $\bra V \ket$, respectively, and the mean values of the Jacobi coordinate $R$, for the ground state wavefunction. Scattering parameters listed are the scattering length a$_s$, the effective range r$_{eff}$, and the value of the cross-section at zero energy $\sigma(E=0)$. As in other works, we have made calculations, only for PES heh2c, for both isotopomers $^3$HeH$_2$ and $^4$HeH$_2$. The last two rows of Table \ref{tab1} show the correction to the ground state energy and to the scattering length obtained by using the two-channel wavefunction of eq. \refeq{trial2}.\\
 The three PESs used show good agreement with each other. The biggest difference, as expected, is in the energy of the bound state, where differences are up to a maximum of 20 \%, whereas differences in the scattering lenght are less than 5\%.  The only vibrational bound state supported shows a highly correlated structure, reflecting its halo character. Its binding energy is given by a big cancellation between the kinetic and potential energies, and both its size and the scattering length are larger than the range of the He-H$_2$ potential. Those characteristics are confirmed by all three surfaces considered. A similar structure, yet further enhanced, is present in $^3$HeH$_2$, which, being lighter, is less bound. The last column of Table \ref{tab1} displays results for $^3$HeH$_2$ calculated with the heh2c PES.\\
Table \ref{tab2} compares the results obtained in this work with the literature. For both $^3$HeH$_2$ and $^4$HeH$_2$ there is a very good agreement with the more accurate calculations of Xiao and Poireir \cite{hexp06}, and of Gianturco {\it et al} \cite{heggl05}, relative to the heh2c PES. From the latter it is shown both the value obtained with a three-dimensional Discrete Variable Representation expansion and the value (indicated as He-2H ) obtained with an approach similar to the one used in this work. Results from Balakrishnan {\it et al.} \cite{hebfd98}, relative to the PES heh2b, also compare quite well with the ones from this work for both He isotopes.\\
For this complex we have also calculated its binding energy and scattering length, relative to PES heh2c, using the more elaborate trial wavefunction of eq. \refeq{trial2}. We obtain the values of $-0.0369$ cm$^{-1}$ for $E$ and $22.648$ for a$_s$, which allow us to conclude that the contribution of neglected terms in the ansatz function \refeq{trial} is negligible.  

\subsection{Ar-H$_2$}

\begin{figure}[h]
\includegraphics[scale=0.50,angle=-90]{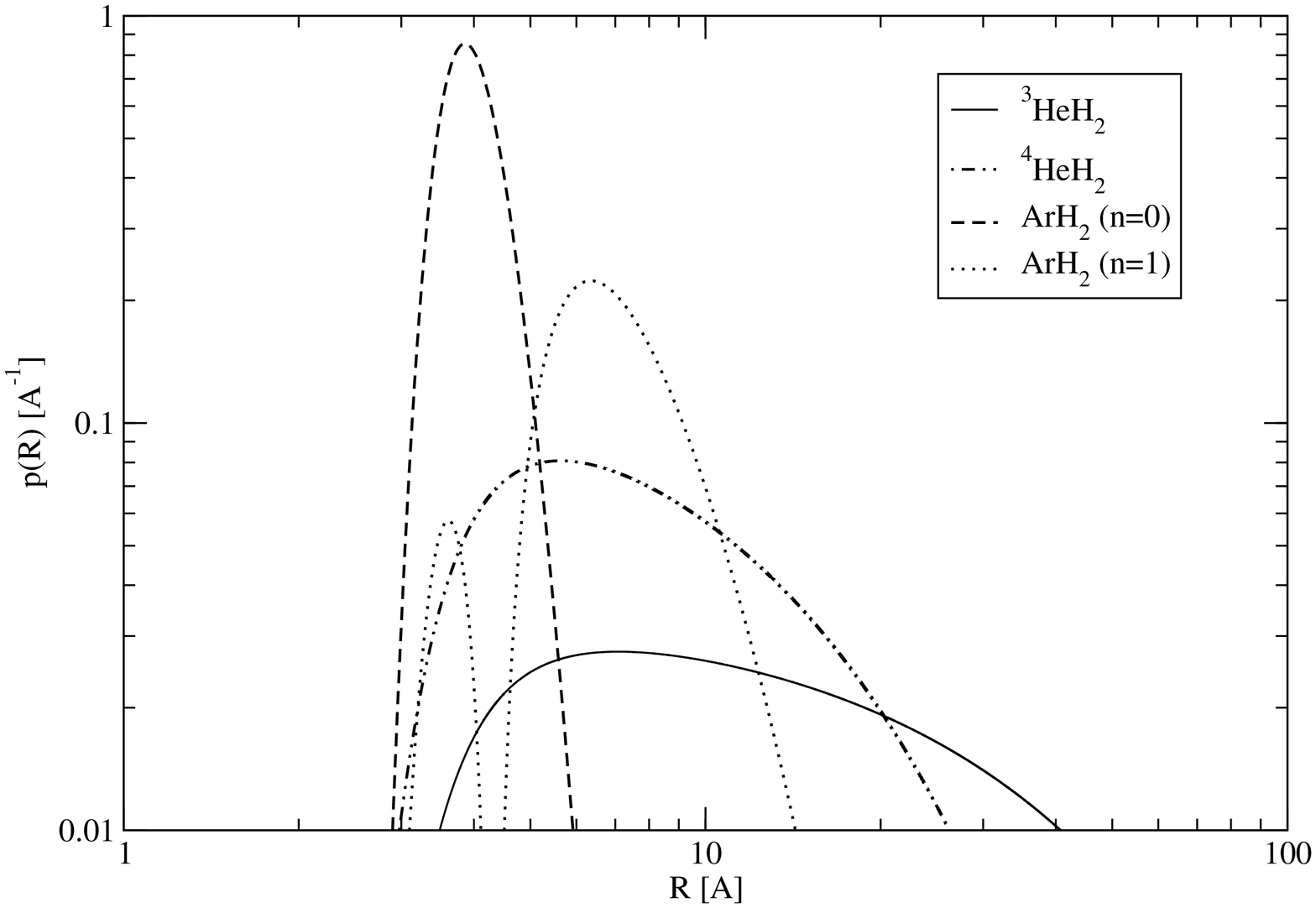}
\caption{}
\label{fig1}
\end{figure}

The attention devoted to this complex in the past years by the quantum chemistry community has resulted in the availability of a very large number of PESs. The most commonly used is the semi-empirical PES of Ref. \cite{arh26} (arh2f). We have also used the surface of Scoles and Rodwell  \cite{nersc82} (arh2s), Le Roy and Hutson (Ref. \cite{xerh86}, arh2b), and Fit 2 of Schwenke {\it et al} \cite{arh22} (arh2c). Table \ref{tab3} lists all relevant parameters for this system. Due to the depth of the effective potential, two vibrational states are present ($n=0,1$). The ground vibrational state ($n=0$) displays a strongly bound character, as for example its binding energy is of the same order of magnitude than its kinetic and potential energies, and the variance of $R$ is much smaller than its mean value. Conversely, the excited vibrational state $n=1$ displays a slightly weaker character, which will be analysed graphically in Fig. \ref{fig1}, altough it can not be considered a halo state. Uudus {\it et al} \cite{arh24} report quantum mechanical calculations of cross sections for ArH$_2$ using the arh2c and arh2f PESs. In particular, their calculated scattering length, which is 9.95 \AA for arh2c and 10.09 \AA for arh2f agree well with the ones in Table \ref{tab3}. All PESs agree within a factor of 10\%. The inclusion of the second channel in the trial wavefunction brings a small contribution to both the binding energies and the scattering length, of less than 1 \%.

\subsection{Ne-H$_2$, Kr-H$_2$ and Xe-H$_2$}

For Ne-H$_2$ there are no recent PESs. We used the surfaces from Ref. \cite{neabh80} (neh2b), from Ref. \cite{neh21} (neh2c), and from Ref. \cite{nersc82} (neh2s). Results for this complex are shown in the first three columns of Table \ref{tab4}. The three surfaces agree within 20\%. They support a single vibrational band, relatively deep in energy (about 20\%) of the potential well. The halo structure present in HeH$_2$ is lost in this system. Tennyson and Sutcliffe \cite{nets82} using PES neh2b obtained an energy of 4.67 cm$^{-1}$ for the vibrational band origin, in agreement with the result of this work. The calculation of accurate surfaces for the heavier atoms, Kr and Xe, is made more difficult by the rapid increase of the number of electrons. For the latter, the only available surface was determined by Le Roy and Hutson \cite{xerh86} (xeh2b). The associated results are shown in the last column of Table \ref{tab4}. For Kr, we used a semi-empirical PES from the same authors (krh2b), a much more recent semi-empirical PES by Wei and co-workers \cite{krh21} (krh2c), and the fully {\it ab initio} surface of Zhou and Xie \cite{krh22} (krh2d). Wei {\it et al} report calculated values for the two band origins, of 28.468 cm$^{-1}$ and 1.653 cm$^{-1}$, which agree with the ones presented in Table \ref{tab4}. Also Zhou and Xie \cite{krh22} report their calculated values for the two band origins, which are 25.688 and 1.048, again in good agreement with the ones calculated  in this work and reported in Table \ref{tab4}. The agreement between PES neh2b and neh2c is rather surprisingly good, with difference of less than 5\% also in quantities, like the scattering length, which are sensitive to every detail of the surface. This agreement leads us to believe that also PES xeh2b would prove reliable if compared to more accurate surfaces. On the other hand, differences between those two PESs (krh2b, krh2c) and krh2d are quite large, and are a consequence of the more shallow depth of the latter. In both KrH$_2$ and XeH$_2$ the two bound states exhibit a deeply bound structure, shown by the mean values of the observables presented in the Table. In fact, the binding energy is comparable to the kinetic and potential energies, the associated radial distrbution is rather peacked around its maximum, and the scattering length is rather small. For all complexes the inclusion of the  second channel changes the results of an  insignificant amount if compared to the uncertainty associated with the PES.

\section{Comparative analysis and Conclusions}

\subsection{Comparative analysis}

Figure \ref{fig1} shows the radial distribution $p(R)$ of the Rg atom with respect to the H$_2$ centre of mass, for selected RgH$_2$ complexes and vibrational states. In particular, the halo states associated with the two HeH$_2$ isotopomers, and the two vibrational states of ArH$_2$ have been plotted. Note the bi-log scale of the plot. The distribution densities associated with the remaining RgH$_2$ complexes display characteristics very similar to the ground vibrational state of ArH$_2$, and are therefore not shown. The ground state of ArH$_2$ displays a rather peaked structure, and the function $p(R)$ is different from zero only in a small range of values of $R$ near its peak. Conversely, the density of the two HeH$_2$ isotopomers show a completely different structure, being more flat and extending over a broad range of values of $R$, well beyond the range of the effective He-H$_2$ potential. Those are typical features of a halo state. The halo structure is more pronounced in $^3$HeH$_2$ than in $^4$HeH$_2$, as expected by the lighter mass of the former \cite{heggl05}. The excited vibrational state of ArH$_2$ shows a structure midway between the two just illustrated: it is quite broad in $R$ but less than the HeH$_2$ complexes. At short distances, all distributions fall rapidly to zero as a consequence of the hard repulsive core of the inter-molecular potential, which is consequence of the closed-shell character of its constituents. At zero incident energy, the largest elastic cross sections are found in He-H$_2$ and Ar-H$_2$, with the others becoming smaller the heavier the Rg atom. This behaviour can be easily interpreted following the analysis of the radial distributions of the bound states, as the two complexes have a halo bound state. It is well known that for halo complexes the scattering length is inversely proportional to the square root of the binding energy ($a_s\approx 1 /\sqrt{E}$), therefore the smaller the latter the bigger the former. Elastic cross sections for a range of energies are shown in Fig. \ref{fig2}. In the limit of zero energy, the difference between the smallest (Xe-H$_2$)  and the largest (He-H$_2$) is more than two orders of magnitude. However, in commenting those results one should always recall that a small change in the PES can result in a very large modification of the cross section for the energy range considered, due to the extreme sensitiveness of the scattering length to the strength of the interaction. In order to have a rough check of the sensitivity of each complex to small modifications of the PES, we have arbitrarily modified each PES by a factor $1+\lambda$, 
\be
V_\lambda(R) = (1+\lambda) V_{eff}(R), 
\label{vla}
\ee
where $\lambda=0$ recovers the real case. This same procedure was applied to test the Efimov character of the excited state of the Helium trimer \cite{esry,pb1}. Figure \ref{fig3} shows the changes in the Rg-H$_2$ scattering length as a function of $\lambda$, compared with the absolute value of the $^4$He-H$_2$ scattering length. The Ne-H$_2$ and Kr-H$_2$ complexes display very little sensitivity to changes in the PES of up to 25\%, in both directions (weakiniing and strengthening), whereas Ar-H$_2$ and Xe-H$_2$ are much more affected by changes to their PESs. In particular, the former will present the same zero energy cross section of $^4$He-H$_2$ when its PES is weakened by a factor $1/5$, whereas the latter when its PES is strengthened by a similar amount. In percentage terms, this implies that a modification of up to 20\% of the Xe-H$_2$ PES can change the associated zero energy cross section of up to two orders of magnitude.

\subsection{Conclusions}

In this paper we have studied the vibrational band origins and ultra-low energy scattering properties of the Rg-H$_2$ complexes, Rg=He, Ne, Ar, Kr and Xe. A simple one-channel trial wavefunction based on a independent H$_2$ model has been used to investigate both the discrete and cold energy range, in conjunction with variational methods. A number of PESs were used to mimic the Rg-H$_2$ inter-molecular interaction. The accuracy of this approach has been checked in a number of ways. Firstly, the contribution of the next  H$_2$ channel was evaluated by implementing a two-channel calculation. This larger calculation was performed for all PESs to calculate the binding energies and scattering lengths for all complexes considered. It showed that the inclusion of the second channel brings a very little contribution, of less than 1 \%, which is insignificant if compared to the uncertainty associated to the molecular PES. Secondly, the results obtained compare well with other published works where more accurate full-dimensional approaches were pursued. The different PESs agree reasonably with each other, in particular the semi-empirical sets of Rodwell and Scoles \cite{nersc82} and Le Roy and Hutson \cite{xerh86} yield results close to the more recent, and presumably more accurate, potentials. The Rg-H$_2$ elastic cross section at ultra-low energy varies by up to two orders of magnitude when the different Rg atoms are considered. He-H$_2$ has the greatest cross section due to the presence of the halo state in the trimer, whereas Xe-H$_2$ has the smallest one. The Ar-H$_2$ elastic cross section is also large, possibly due to the presence of the relatively weakly bound excited vibrational band in this complex. Finally, we have checked the dependence of the atom-dimer scattering length on the strength of the atom-diatom effective potential. It can be argued that the uncertainty in the Rg-H$_2$ PESs can qualitatively affect the conclusions of this study, as the zero-energy scattering length can be dramatically influenced by small (or even by tiny) adjustments of the intermolecular PES. To test those effects, the Rg-H$_2$ PES has been arbitrarily modified by a factor $(1+\lambda)$, and the effect of varying $\lambda$ on the atom-dimer scattering length has been studied. Whereas Ne-H$_2$ and Kr-H$_2$ have shown stability over the range considered ($-0.25\ge \lambda \le 0.25$), Ar-H$_2$ and Xe-H$_2$ proved very sensitive to such changes, and, for instance, the Xe-H$_2$ cross-section increases by two orders of magnitude if the PES is strengthened by $20\%$. As the Xe-H$_2$ PES is also the least well known among the five complexes considered, this result suggests that a more detailed study of the Xe-H$_2$ interaction is needed. \\
 In conclusion, the proposed model can be easily extended to the study of four-body ultra-cold and cold collisions of the type Rg-Rg-H$_2$. In fact, also in this case the H$_2$ molecule can be approximate as a single particle, and the importance of inelastic collisions leading to the formation of Rg-H$_2$ complexes in the four-body collisions can be studied by means of three-body techniques. Regarding applications to Rg-H$_2$ in higher angular momentum states, or to different complexes of experimental interest, such as Rg-NH$_3$, the simple trial wavefunction proposed in this work may be not as effective as in the particular case considered here. Even for those systems, however, it will be interesting to compare results calculated with a many channel and a single channel calculation, recalling that the largest source of error is usually given by the uncertainty in the PES.

\begin{acknowledgements}
I am grateful to R.C.  Forrey for providing the FORTRAN routines relative to the potential arh2c, to T. Gonzalez-Lezana for heh2c, and to D. Xie for krh2d. Routines for arh2f and krh2c were taken from R.J. Le Roy homepage \cite{miscleroy}. I am also grateful to P. Barker, T. Gonzalez-Lezana, A. Kievsky, L. Lodi and J. Tennyson, for useful discussions and comments on this manuscript.
\end{acknowledgements}



\newpage

\begin{table}[h]
\begin{center}
\begin{tabular}{c|rrrrrr} \hline \hline
              &     heh2s\cite{nersc82}  &  heh2b\cite{hemr94}  & \hspace{1cm} &   \multicolumn{2}{c}{heh2c\cite{hebmp03}}    \\
$R_0$         &   2.994  &  3.025  & & \multicolumn{2}{c}{2.993}          \\ 
$R_{min}$     &   3.375  &  3.420  & & \multicolumn{2}{c}{3.373}          \\ 
$\epsilon$    &   9.510  &  8.450  & & \multicolumn{2}{c}{9.499}          \\ \hline
              &$^4$HeH$_2$&$^4$HeH$_2$& & $^4$HeH$_2$&$^3$HeH$_2$ \\
$E$           &   -0.0333 &  -0.0300 & &  -0.0368 &  -0.00348 \\
$\bra K \ket$ &   0.4593 &  0.4269 & &  0.4831 &  0.1564  \\
$\bra V \ket$ &  -0.4926 & -0.4569 & & -0.5198 & -0.1599  \\
$\bra R \ket$ &  26.184  & 14.401  & & 13.372 & 32.922   \\ 
$\rdd $       &  32.076  & 17.711  & & 16.306 & 41.586   \\ \hline
$a_s$         &  23.647  & 24.708  & & 22.690 & 67.638   \\
$r_{eff}$     &   5.08   &  5.16   & &  5.00  &  6.08    \\
$\sigma(E=0)$ [\AA$^2$]& 7027 & 7672 & &  6470 &  57500  \\ \hline
$ \Delta E$      & -0.0002   &  -0.0002 & &  -0.0001& -0.00005\\
$ \Delta a_s$   & -0.069   &  -0.085 & &  -0.052 & -0.478  \\
\hline \hline
\end{tabular}
\caption{List of properties of the HeH$_2$ complex. All distances 
are in \AA\ and energies in cm$^{-1}$. For all potentials the results 
presented refer to the isotopomer $^4$HeH$_2$, whereas for the potential
 heh2c the results are reported for $^4$HeH$_2$ in the left column and for 
$^3$HeH$_2$ in the right column. The upper part of the table reports quantities associated 
to the He-H$_2$ effective potential, namely the crossing point with the $R-$axis $R_0$, 
the minimum $R_{min}$ and the well depth $\epsilon$. The upper middle part reports the 
energy of the bound state $E$, and the mean values of the kinetic energy $K$, the potential 
energy $V$, and the intermolecular distance $R$, along with its square $R^2$.
The lower middle part reports the scattering length a$_s$, the effective range r$_eff$ and 
the zero energy cross section $\sigma(E=0)$. Finally, the bottom part reports the correction 
to the binding energy and to the scattering length obtained by using the two-channel 
trial wavefunction with respect to the single channel one ($\Delta x = x^{(2)}-x^{(1)}$).
 The values used of the atom's masses 
are 1.00794 amu, 4.00260 amu and 3.01603 amu for H, $^4$He and $^3$He, respectively.}
\label{tab1}
\end{center}
\end{table}

\begin{table}[h]
\begin{center}
\begin{tabular}{c|rrrr|rr} \hline \hline
              & \multicolumn{4}{c}{heh2c\cite{hebmp03}} &  \multicolumn{2}{c}{heh2b\cite{hemr94}} \\ \cline{2-5} \cline{6-7}
              & Ref. \cite{hexp06}& \multicolumn{2}{c}{Ref. \cite{heggl05}} & this work & Ref. \cite{hebfd98} & this work \\
              &         &He-H$_2$ &  He-2H  &          &        &         \\ \cline{3-4}
 $^4$He-H$_2$ & -0.03640 & -0.03634 & -0.03803 &  -0.0368  & -0.0298 & -0.0300  \\
 $^3$He-H$_2$ & -0.00327 & -0.002916 &  -      &  -0.00348 & -0.0016 & -0.0019 \\
\hline \hline
\end{tabular}
\caption{Compilation of the values for the binding energies, in cm$^{-1}$, for $^4$HeH$_2$ and $^3$HeH$_2$ present in the literature, compared to the values obtained in this work. The agreement is good for both isotopomers and for the two PESs considered.}
\label{tab2}
\end{center}
\end{table}

\begin{table}[h]
\begin{center}
\begin{tabular}{c|rrrr} \hline \hline
              &     arh2s\cite{nersc82}  &  arh2b\cite{xerh86}  &    arh2c\cite{arh22} &  arh2f\cite{arh25} \\
$R_0$         &     3.144  &   3.179 & 	3.194 &   3.179 \\ 
$R_{min}$     &     3.551  &   3.580 &	3.610 &   3.591 \\ 
$\epsilon$    &    53.922  &  50.88  & 50.982 &  50.479 \\ \hline
              & \multicolumn{4}{c}{n=0} \\ \hline
$E$         &    -23.952  &  -22.113 & -22.205 &  -21.975 \\
$\bra K \ket$ &    11.712  &  11.156 & 11.212 &  11.078 \\
$\bra V \ket$ &   -35.664  & -33.269 &-33.417 & -33.053 \\
$\bra R \ket$ &     3.989  &   4.046 &  4.053 &   4.046 \\
$\rdd $       &     4.020  &   4.078 &  4.085 &   4.078 \\ \hline
              & \multicolumn{4}{c}{n=1} \\ \hline
$E$         &     -0.617  &   -0.478 &  -0.395 &   -0.416 \\
$\bra K \ket$ &     3.688  &   3.142 &  2.810 &   2.911 \\
$\bra V \ket$ &    -4.305  &  -3.621 & -3.205 &  -3.327 \\
$\bra R \ket$ &     7.348  &   7.752 &  8.029 &   7.940 \\
$\rdd $       &     7.680  &   8.140 &  8.466 &   8.364 \\\hline
$a_s$         &     8.710  &   9.523 & 10.102 &   9.930 \\
$r_{eff}$     &     6.12   &   5.22  &  5.22  &   5.21  \\
$\sigma(E=0)$ [\AA$^2$] &     953 & 1140 &1282 & 1239 \\ \hline
$E_1^{(2)}$   &   -0.016   &  -0.016 &  -0.015&   -0.015\\ 
$E_2^{(2)}$   &   -0.007   &  -0.002 &  -0.001&   -0.001\\ 
$a_s^{(2)}$   &   -0.002   &  -0.009 &  -0.010 &  -0.009 \\ 
\hline \hline
\end{tabular}
\caption{List of properties of the ArH$_2$ complex. All distances 
are in \AA\ and energies in cm$^{-1}$. The mass of Ar has been set as 39.948 amu.
 Results are shown for the ground (n=0) and excited (n=1) vibrational states.
 For the explanation of the 
different reported quantities see caption of Table \ref{tab1}.}
\label{tab3}
\end{center}
\end{table}

\begin{table}[h]
\begin{center}
\begin{tabular}{c|rrrrrrrrr} \hline \hline
              & \multicolumn{3}{c}{Ne-H$_2$}  &  \hspace{1cm} &  \multicolumn{3}{c}{Kr-H$_2$} & \hspace{1cm} & \multicolumn{1}{c}{Xe-H$_2$}\\
              &  neh2b\cite{neabh80}  &  neh2c\cite{neh21} & neh2s\cite{nersc82} &  &  krh2b\cite{xerh86} & krh2c\cite{krh21} & krh2d\cite{krh22} &  & xeh2b\cite{xerh86} \\
$R_0$         &   2.920 &  2.963 &  2.924 & &   3.304 &  3.304 &  3.331 & &   3.510 \\ 
$R_{min}$     &   3.300 &  3.342 &  3.300 & &   3.720 &  3.730 &  3.759 & &   3.940 \\ 
$\epsilon$    &  22.978 & 21.071 & 23.565 & &  58.849 & 58.466 & 54.517 & &  64.914 \\ \hline
              & \multicolumn{9}{c}{n=0} \\ \hline		
$E$           &   -4.741 &  -3.959 &  -4.851 & &  -28.453 & -28.493 & -25.648 & & -33.194 \\ 
$\bra K \ket$ &   5.448 &  4.917 &  5.563 & &  12.203 & 12.090 & 11.518 & &  12.949 \\
$\bra V \ket$ & -10.189 & -8.876 & 10.414 & & -40.657 &-40.583 &-37.166 & & -46.142 \\
$\bra R \ket$ &   4.248 &  4.379 &  4.236 & &   4.127 &  4.128 &  4.173 & &   4.320 \\
$\rdd $       &   4.340 &  4.481 &  4.326 & &   4.154 &  4.155 &  4.202 & &   4.344 \\ \hline
              & \multicolumn{9}{c}{n=1} \\ \hline		
$E$           &         &        &        & &   -1.734 &  -1.660 &  -1.074 & &   -3.139 \\
$\bra K \ket$ &         &        &        & &   6.529 &  6.423 &  4.982 & &   8.996 \\
$\bra V \ket$ &         &        &        & &  -8.263 & -8.083 & -6.056 & & -12.135 \\
$\bra R \ket$ &         &        &        & &   6.409 &  6.434 &  6.890 & &   6.153 \\
$\rdd $       &         &        &        & &   6.585 &  6.614 &  7.120 & &  6.278  \\ \hline
$a_s$         &   3.360 &  3.846 &  3.309 & &   5.509 &  5.659 &  6.961 & &   1.819 \\
$r_{eff}$     &  14.37  & 10.70  &  14.92 & &  14.83  & 13.94  &  9.65  & & 288.78  \\
$\sigma(E=0)$ [\AA$^2$]&  142 & 186  &  138 & & 381 & 402 & 609 & & 41.6 \\ \hline
$E_1^{(2)}$   &   -0.006& -0.006 & -0.007 & &  -0.024 &  -0.021& -0.024 &  &   -0.030 \\ 
$E_2^{(2)}$   &         &        &        & &  -0.003 &  -0.002& -0.003 &  &   -0.005 \\ 
$a_s^{(2)}$   &   -0.004& -0.003 & -0.003 & &  -0.006 &  -0.006& -0.007 &  &   -0.012 \\ 
\hline \hline
\end{tabular}
\caption{List of properties of the XH$_2$ complexes, with X= Ne,  Kr, Xe. All distances 
are in \AA\ and energies in cm$^{-1}$. The values used for the atom's masses are 20.1797 amu, 
83.8 amu and 131.293 amu for Ne, Kr and Xe, respectively.
 Results are shown for the ground (n=0) and excited (n=1) vibrational state.
 For the explanation of the 
different reported quantities see caption of Table \ref{tab1}.}
\label{tab4}
\end{center}
\end{table}

\clearpage
\newpage 

\noindent Figure Captions

\noindent Figure 1. 

Density distribution $P(R)$ for selected RgH$_2$ vibrational states. In particular, the density distributions for the bound states of $^3$HeH$_2$ and $^4$HeH$_2$, and the two vibrational states for ArH$_2$ are plotted. The densities shown refer to calculations made using the heh2c and arh2b potentials. Note the bi-log scale. 

\noindent Figure 2. 

Elastic cross sections for the five Rg-H$_2$ complexes, plus the $^3$He-H$_2$ isotopomer. The energy is given with respect to the centre of mass frame, ranging from 10$^{-3}$ to 10 cm$^{-1}$. Note the bi-log scale. The curves shown are: dotted line, dot-dash line, dot-dot-dash line, dash-dash-dot line, dashed line and continuum line for $^3$He-H$_2$, $43$He-H$_2$, Ar-H$_2$, Kr-H$_2$, Ne-H$_2$ and Xe-H$_2$, respectively. The displayed cross sections are relative to phase-shifts obtained with the heh2c, neh2s, arh2b, krh2b and xeh2b potentials.

\noindent Figure 3. 

\begin{figure}[h]
\includegraphics[scale=0.50,angle=-90]{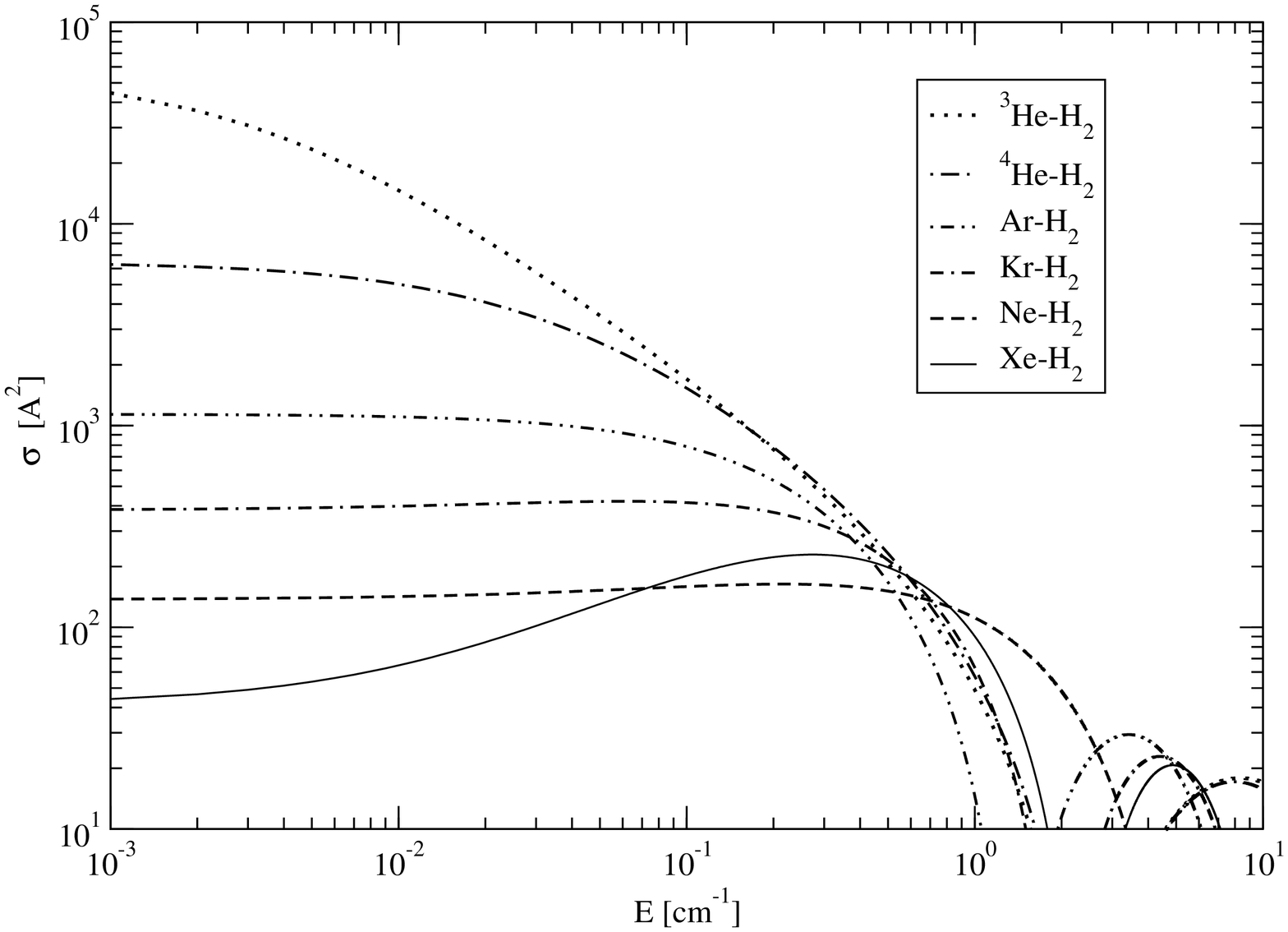}
\caption{}
\label{fig2}
\end{figure}

\begin{figure}[h]
\includegraphics[scale=0.50,angle=-90]{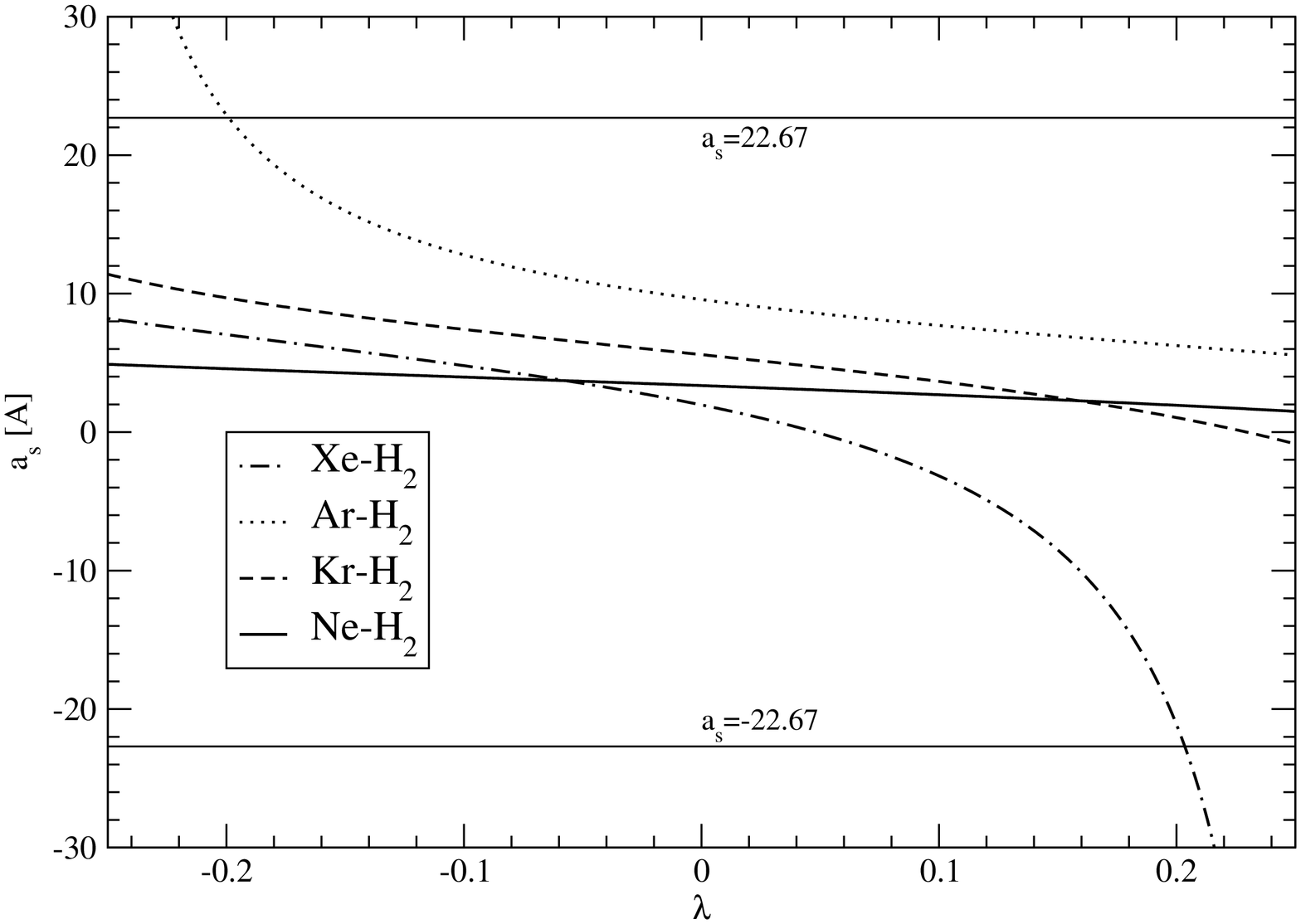}
\caption{}
\label{fig3}
\end{figure}

Dependence of the scattering length a$_s$ on the parameter $\lambda$ (eq. \refeq{vla}). The two horizontal continuum lines indicate the threshold where a$_s$ reaches the same value in absolute terms of the $^4$He-H$_2$ scattering length. Calculations have been made using the same potentials as indicated in the caption of Fig. \ref{fig2}.

\newpage

\newpage

\newpage

\end{document}